\documentclass{article}
\usepackage{spconf,amsmath,graphicx,hyperref}
\usepackage{enumitem}
\usepackage{graphicx}
\usepackage{pifont}
\usepackage{adjustbox}
\newcommand{\cmark}{\ding{51}}%
\newcommand{\xmark}{\ding{55}}%
\usepackage{amssymb}
\usepackage{xspace}

\title{MaskVCT: Masked Voice Codec Transformer for Zero-Shot Voice Conversion With Increased Controllability via Multiple Guidances}
\name{\begin{tabular}{c}
Junhyeok Lee, Helin Wang, Yaohan Guan, Thomas Thebaud, \\
Laureano Moro-Velazquez, Jes\'us Villalba, Najim Dehak
\end{tabular}}

\address{Center for Language and Speech Processing, Johns Hopkins University, Baltimore, MD, USA}

\begin{document}
\ninept
\maketitle
\begin{abstract}
We introduce MaskVCT, a zero‑shot voice conversion (VC) model that offers multi‑factor controllability through multiple classifier‑free guidances (CFGs). While previous VC models rely on a fixed conditioning scheme, MaskVCT integrates diverse conditions in a single model. To further enhance robustness and control, the model can leverage continuous or quantized linguistic features to enhance intelligibility and speaker similarity, and can use or omit pitch contour to control prosody. These choices allow users to seamlessly balance speaker identity, linguistic content, and prosodic factors in a zero‑shot VC setting. Extensive experiments demonstrate that MaskVCT achieves the best target speaker and accent similarities while obtaining competitive word and character error rates compared to existing baselines. Audio samples are available at \url{https://maskvct.github.io/}.
\end{abstract}
\vspace{-4pt}
\section{Introduction}
\vspace{-4pt}
Voice conversion (VC) aims to transform the identity of a source utterance to match a target speaker while preserving the original linguistic content. A fundamental challenge in this task is the effective disentanglement of linguistic content from pitch and timbre.
Recent VC models \cite{resynthesis,nansy,nansypp,freevc, diffhiervc,maskgct} have increasingly adopted representation from pre-trained self-supervised learning models as linguistic features.
However, despite being described as ``disentangled", these features often remain ``leaky", retaining sufficient pitch and speaker information to reconstruct the original mel-spectrogram \cite{nansy}.

This information leakage creates a divide in model conditioning. 
\textit{Pitch-conditioned} models \cite{nansy,nansypp,diffhiervc} offer high controllability via frame-wise pitch guidance, but rely on accurate external pitch predictors.
In contrast, \textit{pitch-unconditioned} models \cite{freevc, maskgct,genvc} generate speech without explicit pitch conditioning.
These models rely on the pitch information that remains entangled within linguistic features; this entanglement notably enables the generation of tonal languages like Chinese \cite{maskgct}.

\begin{figure*}
    \centering
    \includegraphics[width=0.99\linewidth]{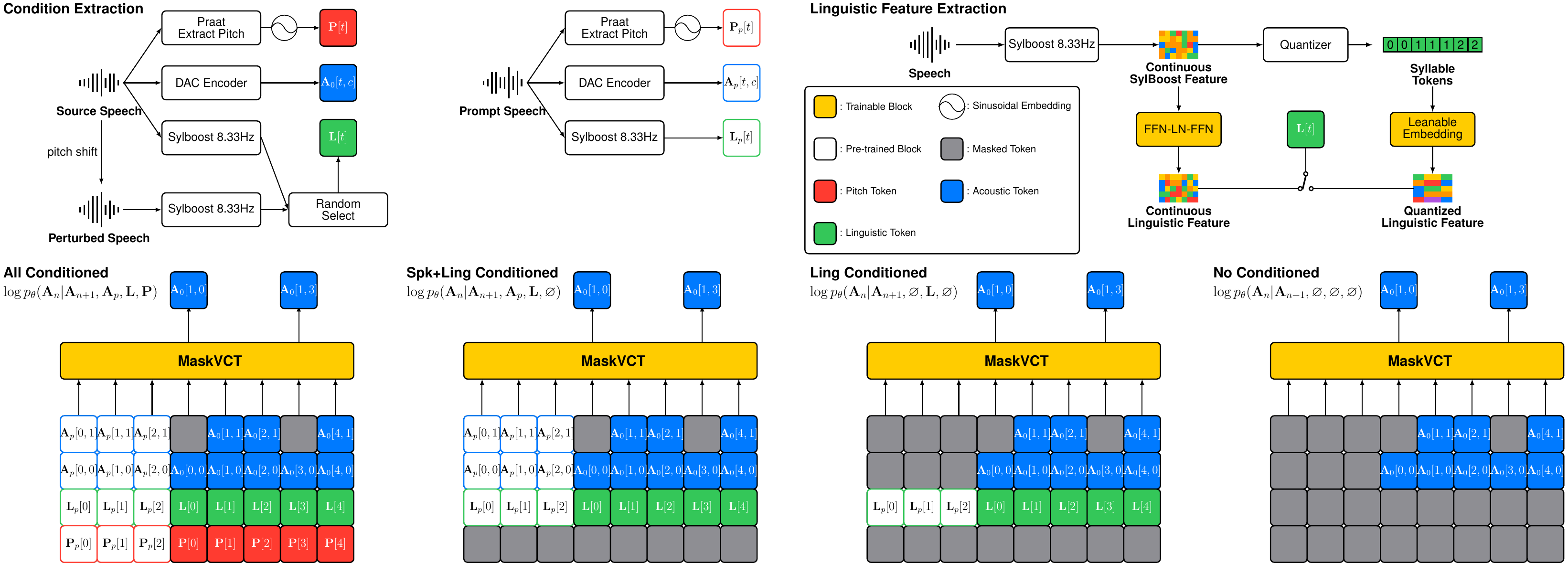}
    \vspace{-10.0pt}
    \caption{Overall system description of MaskVCT. 
    We perform column-wise addition of the embeddings and feed the result into MaskVCT.
    We employ 9 codebooks for DAC, but display only 2 here for brevity. All models operate at 50 Hz frame rate.}
    \vspace{-12pt}
    \label{fig:overall}
\end{figure*}

Recently, \textit{zero-shot} models have emerged, allowing conversion to previously unseen speaker voices without additional fine-tuning \cite{nansy,freevc,diffhiervc,maskgct}.
Specifically, codec-based language models (CodecLMs)
have become a dominant paradigm. 
By treating discretized neural audio tokens as targets \cite{dac,naturalspeech3}, CodecLMs formulate speech generation as a Transformer-based classification task \cite{transformer}, 
utilizing either autoregressive  \cite{valle} or masked modeling \cite{maskgct} conditioned on short speaker prompts.
Despite their popularity, however, we identify a critical limitation that zero-shot VC with prompting also struggles to discard source pitch information. This information leakage results in converted speech that retains the source's intonation rather than adopting the target's style.

To address these limitations of controllability and disentanglement, we propose \textbf{MaskVCT}, a masked generative model for zero-shot voice conversion that combines temporal-quantized syllabic features with multiple classifier-free guidances to enable enhanced controllability. 
By using syllabic representations from SylBoost \cite{syllablelm}, MaskVCT gains flexibility in pitch modulation, phoneme selection, and accent variation within each segment. 
By leveraging multiple classifier-free guidances (CFGs) \cite{cfg}, we allow for a dynamic balance between conditions at inference time, effectively creating a highly controllable system.
Unlike prior VC models bound to a single condition, MaskVCT supports different conditions and inference modes in a \textbf{single model}, with respect to:
\vspace{-0pt}
\begin{itemize}[leftmargin=*]
  \item {Linguistic representation (categorical):} discrete vs.\ continuous.
  \item {Pitch (categorical):} source-conditioned vs.\ target-driven contour.
  \item {CFG weights (continuous):} balancing content, speaker, and pitch.
  \vspace{-6pt}
\end{itemize}
From those potential configurations, we propose two modes: MaskVCT-All, which prioritizes pitch following and intelligibility, and MaskVCT-Spk, which focuses on speaker similarity.
Both modes deliver strong performance across all metrics, especially MaskVCT-Spk achieves the highest target speaker and accent similarity.

\vspace{-2pt}
\section{MaskVCT}
\vspace{-2pt}
\subsection{Linguistic Conditioning}

Self-supervised speech representations provide time-aligned and fast-varying phonetic features \cite{ss_phonetic}.
Therefore, VC models built on them tend to reproduce source-aligned speech. 
In our preliminary study, even pitch-unconditioned and linguistics-only-conditioned models \cite{freevc,maskgct} retain high pitch correlation with the source, indicating pitch leakage.
This motivates a coarser and pitch-stripped representation for source-agnostic pitch generation.

Recent works propose syllabic representations \cite{syllablelm,sylber} that discretize speech into slow-varying units by quantizing both within-vector and across adjacent frames.
Also, they relax strict source alignment to enable phoneme and accent variations.
We adopt these temporally coarse-grained tokens to mitigate non-linguistic attribute leakage.
We use the discrete tokens from the pre-trained syllabic model SylBoost and its quantizer \cite{syllablelm}, and train learnable embeddings for each token.

Continuous and discrete linguistic representations offer complementary strengths: continuous features improve intelligibility and phoneme alignment, whereas discrete syllable tokens better preserve target timbre/speaker similarity \cite{sylber} and mitigate source speaker traits leakage; conversely, continuous features can weaken speaker fidelity \cite{vevo}. 
To expose this trade-off, we implement dual conditioning paths: (i) discretized syllabic tokens and (ii) continuous features via a lightweight FFN–LayerNorm–FFN projection. We train with balanced sampling as 50\% for each, allowing the model to run in either mode, so users can choose intelligibility-forward or target timbre-faithful at inference.

\subsection{Pitch Conditioning}

We map continuous pitch to a vector using sinusoidal embedding \cite{transformer}, making the system agnostic to pitch-extractor resolution and requiring no fine-tuning when changing extractors. 
Unlike the standard interleaved layout for positional embedding, we concatenate sine and cosine terms to avoid confounding multi-head attention.
As the human perception of frequency is related to the log-scale, we convert pitch to the log-scale, adding an epsilon value of 1 Hz to the frequency, to prevent negative infinity with fundamental frequencies of 0 Hz. 
Therefore, for the frequency $\mathbf{f}\in\mathbb{R}^T$, the pitch embedding $\mathbf{P}\in\mathbb{R}^{T\times d}$ is defined as follows:
\vspace{-5pt}
\begin{equation}
    \mathbf{P}[t,i] =\begin{cases}
    \sin\left(\frac{\log(1+\mathbf{f}[t])}{10000^{2\cdot i/d}}\right) & \text{if $i<d/2$} \\
    \cos\left(\frac{\log(1+\mathbf{f}[t])}{10000^{2\cdot (i-d/2)/d}}\right)& \text{if $i\geq d/2$}
    \end{cases} 
\end{equation}
where $i$ in 0 to $d-1$, $t$ in 0 to $T-1$, and $d$ is the model's dimension. 
We extract all pitch contours with a rate of 50 Hz using Praat \cite{praat}.
\subsection{Speaker Conditioning}

To perform zero-shot VC, we implemented a speaker prompt mechanism similar to Vall-E \cite{valle}. 
We process 3-second of the reference speech to extract the \textit{prompt} conditions and then concatenate them to the beginning of the source conditions. 
By this, we can leverage in-context learning by showing how the given speaker reads the given conditions to increase the quality and condition satisfaction.

\subsection{Masked Codec Language Model}

Inspired by masked language modeling \cite{bert},
MaskGIT \cite{maskgit} frames image synthesis as discrete token unmasking.
Extending this idea, similar to CodecLMs \cite{valle,musicgen}, non-autoregressive codec-based models adopt masked modeling to speech \cite{maskgct,soundstorm},
typically on residual vector quantization (RVQ) codecs \cite{dac,encodec}.
Recently, MaskGCT \cite{maskgct} achieved a state-of-the-art zero-shot text-to-speech with a masked model, and its S2A module can be utilized as a VC model. 

The masked model is trained by reconstructing the source's encoded acoustic tokens 
$\mathbf{A}_0 \in \{1,2,\dotsc,K\}^{T\times Q}$ from a masked input $\mathbf{A}_{u,q} =\mathbf{A}_0 \odot \mathbf{M}_{u,q}$, where $K$ is the codebook vocabulary size, $T$ the number of frames, $Q$ the number of the RVQ codebooks, $q\in\{0,1,\dotsc,Q-1\}$ a target RVQ codebook layer, $u\in [0,1]$ a masking timestep, and $\mathbf{M}_{u,q}\in\{0,1\}^{T \times Q}$ a binary mask. 
To be specific, for the mask $\mathbf{M}_{u,q}$, all layers below $q$ remain unmasked, all layers above $q$ are fully masked, and only layer $q$ is masked according to a masking schedule.
The reconstruction process can be framed as a classification task where the model $\theta$ learns the conditional probability $p_\theta(\mathbf{A}_{0}|\mathbf{A}_{u,q},\mathbf{C})$, where $\mathbf{C}$ are the conditioning features.
During training, we randomly sample a masking time step $u\sim\mathcal{U}[0,1]$, a target codebook layer $q\sim p(q) = 1 -\frac{2(q+1)}{Q(Q+1)}$ \cite{maskgct}, and cosine scheduled mask $\mathbf{M}_{u,q}\sim\mathrm{Bernoulli}\left(\cos\left(\pi u/2\right)\right)
$ \cite{maskgit}.
We train our model using a simple classification loss applied exclusively to the masked tokens, defined as follows:
\vspace{-4pt}
\begin{equation}
    \mathcal{L}_\text{mask} = \displaystyle \mathop{\mathbb{E}}_{t | \mathbf{M}_{u,q}[t,c]=0} \left[-\log p_{\theta} \left(\mathbf{A}_{0}[t,c]|\mathbf{A}_{u,q},\mathbf{C} \right) \right].
\end{equation}
\vspace{-7pt}

During inference, we start from all-masked state $\mathbf{A}_{1,0}=\mathbf{A}_N$, and iteratively unmask tokens over $N$ iterations, producing intermediate states $\mathbf{A}_n$, until reaching the fully unmasked state $\mathbf{A}_0$, where $n \in \{0,1,\dotsc,N\}$. 
Each iteration index $n$ corresponds to a codebook index $q$ according to a predefined schedule.
The model estimates logits for masked positions as $\log p_\theta(\mathbf{A}_n \mid \mathbf{A}_{n+1}, \mathbf{C})$.

\vspace{-1pt}
\subsection{Multiple Classifier-Free Guidances}
\vspace{-1pt}

CFG \cite{cfg} is a commonly used technique to improve conditional generation at inference time and is also used in CodecLMs that require fine time-aligned conditioning for accurate acoustic reconstruction \cite{valle,maskgct}.
There are several studies that applied dual CFG \cite{voiceldm,dualspeech},
notably, DualSpeech \cite{dualspeech} uses dual CFG to jointly steer intelligibility and speaker fidelity via phoneme and speaker-aware phoneme embeddings.

In VC, where outputs must simultaneously satisfy multiple conditioning factors, including speaker identity, linguistic content, and pitch contour, we extend CFG to triple CFGs.
Within masked generative modeling framework \cite{maskgct,maskgit,soundstorm}, we define speech synthesis as a conditional probability with three factors as speaker prompt $\mathbf{A}_p$, linguistic content $\mathbf{L}$, and pitch $\mathbf{P}$ as $\log p_{\theta}(\mathbf{A}_{n}|\mathbf{A}_{n+1},
\mathbf{C})$, where the conditioning variable $\mathbf{C}$ is set to either $(\mathbf{A}_{p}, \mathbf{L}, \mathbf{P})$ or $(\mathbf{A}_{p}, \mathbf{L}, \varnothing)$ depending on pitch condition.
Define the linguistic-only path as
$\log p_\theta(\mathbf{A}_n\mid \mathbf{A}_{n+1},\mathbf{L}) :=\log p_\theta(\mathbf{A}_n\mid \mathbf{A}_{n+1},\varnothing,\mathbf{L},\varnothing)$.
Following DualSpeech \cite{dualspeech}, the speaker condition couples the prompt with linguistic as
$\log p_\theta(\mathbf{A}_n\mid \mathbf{A}_{n+1},\mathbf{A}_p,\mathbf{L},\varnothing)$.
Since pitch alone is ill-posed, we use it jointly with speaker and text as
$\log p_\theta(\mathbf{A}_n\mid \mathbf{A}_{n+1},\mathbf{A}_p,\mathbf{L},\mathbf{P})$.

Unlike other generative modeling tasks, VC is a strictly conditional generative task that explicitly conditions on and preserves the source linguistic content. 
Therefore, we adopt a modified CFG formulation; instead of subtracting the unconditioned logit $\log p_\theta(\mathbf{A}_{n}|\mathbf{A}_{n+1}):=\log p_\theta(\mathbf{A}_{n}|\mathbf{A}_{n+1}, \varnothing,\varnothing,\varnothing)$ from the conditioned logit, we subtract the linguistic-only conditioned logit from the detailed conditioned logit as follows:
\vspace{-5pt}
\begin{align}
\label{eq:cfg}
     & \log \Tilde{p}_\theta(\mathbf{A}_{n}|\mathbf{A}_{n+1}, \mathbf{C})    \nonumber\\
        &=\log p_\theta(\mathbf{A}_{n}|\mathbf{A}_{n+1},\mathbf{L}) \nonumber\\
        &+\omega_{\text{all}}\left(\log p_\theta(\mathbf{A}_{n}|\mathbf{A}_{n+1}, \mathbf{A}_p,\mathbf{L},\mathbf{P})-\log p_\theta(\mathbf{A}_{n}|\mathbf{A}_{n+1},\mathbf{L}) \right) \nonumber \\
        &+\omega_{\text{spk}}\left(\log p_\theta(\mathbf{A}_{n}|\mathbf{A}_{n+1}, \mathbf{A}_p,\mathbf{L},\varnothing)-\log p_\theta(\mathbf{A}_{n}|\mathbf{A}_{n+1},\mathbf{L}) \right) \nonumber \\
        &+\omega_{\text{ling}}\left(\log p_\theta(\mathbf{A}_{n}|\mathbf{A}_{n+1}, \mathbf{L})-\log p_\theta(\mathbf{A}_{n}|\mathbf{A}_{n+1}) \right),
\end{align}
where $\omega_\text{all}$, $\omega_\text{spk}$, $\omega_\text{ling}$ are CFG coefficients of each conditioning, respectively.
We trained a model with four different conditions, as shown in Figure \ref{fig:overall}, and all-conditioned, spk-conditioned, ling-conditioned, and null-conditioned are randomly sampled in a 6:1:2:1 ratio.
For each codebook layer of audio tokens, linguistic and pitch conditions have their own mask tokens, which replace the original token to mask it.

\subsection{Model Architecture Details}

We utilize a simple Transformer encoder-based architecture \cite{transformer}, especially PreLN with rotary positional embedding.
For simplicity, FFNs use a ReLU activation and LayerNorm.
After the 16 Transformer encoder layers, each with 16 attention heads, model dimension of 1024, FFN dimension of 4096, we attach separate classification heads for each codebook index.

\section{Experiments}

\subsection{Baselines}

For our baseline, we utilized the official implementation and checkpoint of Diff-HierVC\footnote{\url{https://github.com/hayeong0/Diff-HierVC}} \cite{diffhiervc}, FACodec\footnote{\url{https://hf.co/amphion/naturalspeech3_facodec}} \cite{naturalspeech3,amphion}, MaskGCT\footnote{\url{https://hf.co/amphion/MaskGCT}} \cite{maskgct},  FreeVC\footnote{\url{https://github.com/OlaWod/FreeVC}} \cite{freevc}, GenVC\footnote{\url{https://github.com/caizexin/GenVC}}~\cite{genvc}.
For accent conversion, we only tested the models that achieved the best objective metrics in the main evaluation.
Table \ref{tab:baselines} shows the number of parameters and training dataset hours for each model. 
\begin{table}[h]
    \centering
    \resizebox{0.85\linewidth}{!}{%
    \begin{tabular}{l|c c}
    \hline
    Model & \# Parameter & Total Hours \\
    \hline
    Diff-HierVC & 18M (300M+13M) & 245 (436K)\\
    FACodec & 140M & 60K\\ 
    MaskGCT-S2A & 335M (600M+214M) & 100K (4.5M)\\
    FreeVC & 39M (317M) & 44 (94K)\\
    GenVC & 484M (94M+3M) & 47K (960)\\
    \hline
    \bf{MaskVCT}& 234M (94M+74M) & 100K (960)\\
    \hline 
    \end{tabular}
    }
        \vspace{-8pt}
    \caption{Comparison of baseline models and ours with number of parameters and training hours.
    The parentheses next to parameters indicate the number of parameters pre-trained models they utilized.
    The parentheses next to the hour are the total hours spent training the linguistic feature extractor.}
    \label{tab:baselines}
        \vspace{-8pt}
\end{table}

\subsection{Training}
We employ the official checkpoint of the DAC 16 kHz \cite{dac} as a codec, from which we extract 9 codebook indices to encode each audio frame. For syllabic representations, we use SylBoost 8.33 Hz \cite{syllablelm}, which has a minimum vocabulary size.
To improve robustness to phase-related inconsistency \cite{inconsistency}, we apply PhaseAug \cite{phaseaug} to all inputs before DAC encoder. 
Although SylBoost yields consistent features, self-reconstruction can leak information \cite{nansy}; thus, following NANSY \cite{nansy}, we pitch-shift clean speech to create perturbed variants and train with a 50\% property for each case.

We trained our model from scratch for 250k steps on 2 A100 GPUs using AdamW with a batch size of 168 and a learning rate of 0.0002.
Additionally, both the layer drop rate and the dropout rate were set to 5\%.
We apply SpecAugment \cite{specaugment} to combined embeddings by randomly masking 10\% of the channel dimension of the input vectors.
For all cases, we utilize 3-second speaker prompts and 10.24-second as the source.

\subsection{Dataset}
Our model was trained on diverse corpora to enhance its generalization across accents and speaking styles. We utilized the train-clean subsets of LibriTTS‑R \cite{librittsr}, the train subset of MLS‑en \cite{mls}, VCTK \cite{vctk}, LibriHeavy-Large \cite{librilight,libriheavy}, the clean subset of HiFi‑TTS \cite{hifitts}, LJSpeech \cite{ljspeech}, and the speech subset of RAVDESS \cite{ravdess}.

We utilized the test-clean subset of LibriTTS-R \cite{librittsr} to test the general performance of VC models. 
Since the models were not trained with the long length, we select the utterances that are shorter than 10 seconds and slice 3 seconds for the prompt speeches.
We test the models with 511 sample pairs.

To assess accent conversion performance, we utilize the non-native English speech corpus L2-ARCTIC \cite{l2arctic}. 
Two conversion directions are considered:
{Libri$\rightarrow$L2}: converting from near-native LibriTTS-R speech to accented speech, and {L2$\rightarrow$L2}: conversion between distinct accents.
Test utterance pairs were selected using the same criteria as in the main test.
Additionally, while the ground truth (GT) exhibits low intelligibility, we omit intelligibility metrics for this test.

\begin{table*}[th]
    \centering
    \resizebox{0.94\linewidth}{!}{%
    \begin{tabular}{l|c|cc|cc|ccc|cc}
    \hline
     Model        
     & NFE 
     & \# Token 
     & Pitch
     & WER ($\downarrow$)  
     & CER ($\downarrow$) 
     & S-SIM ($\uparrow$)  
     & FPC ($\updownarrow$) 
     & SS-MOS ($\uparrow$)
     & UTMOS ($\uparrow$) 
     & Q-MOS ($\uparrow$) 
       \\
    \hline
    Ground Truth (GT)              &&&& \bf{2.95}      &  \bf{1.44}  & \bf{0.890} & \bf{1.000} 
    & \bf{3.99$\pm$0.21} 
    & \bf{3.25$\pm$0.04} 
    & \bf{3.93$\pm$0.23} 
    
    \\      
    GT - DAC reconstructed &1&&&   3.14   & 1.50   & 0.886 & 0.760 
    & $-$
    & 3.09$\pm$0.03 
    & 4.02$\pm$0.21 
     \\      
    \hline
    Diff-HierVC  & 30+6 & $\infty$ & \cmark & 5.31      &   2.62  & 0.865  & 0.388 
    & 3.30$\pm$0.30
    & 2.97$\pm$0.04
    & 2.83$\pm$0.24 \\
    FACodec  & 1 & 1024$^2$ & \cmark   & \bf{3.55}      &   \bf{1.66}  & \underline{0.883}  & 0.360 
    & 3.30$\pm$0.23  
    & 3.02$\pm$0.04 
    & 2.90$\pm$0.24 \\
    MaskGCT-S2A  & 66 & 8192 & \xmark     & 5.18       &   2.89  & 0.863 & 0.396 
    & 3.02$\pm$0.28 
    & \bf{3.24$\pm$0.03}  
    & \bf{4.02$\pm$0.21}\\
    FreeVC & 1 & $\infty$ &  \xmark & \underline{3.96} & \underline{1.91} & 0.855 & \bf{0.420} 
    & \underline{3.47$\pm$0.23} 
    & 2.79$\pm$0.03 
    & \underline{3.55$\pm$0.26}\\
    GenVC  & 2$T$ & 256 & \xmark & 7.18 & 3.57 & 0.846 & 0.192 
    & 3.03$\pm$0.27 
    & 2.49$\pm$0.04 
    & 2.51$\pm$0.23 \\
    \hline
    \bf{MaskVCT-All}        &64& $\infty$  & \cmark   & {4.68} & {2.22}  & {0.865} & \underline{0.417} 
    & 2.59$\pm$0.24
    & {3.05$\pm$0.03} 
    & \underline{3.54$\pm$0.24} \\
    \bf{MaskVCT-Spk}             &64& 2048   & \xmark  & 6.47 & 3.09 & \bf{0.895} & \textit{0.167}  
    & \bf{3.69$\pm$0.26}
    & \bf{3.17$\pm$0.03} 
    & 3.44$\pm$0.22\\
    \hline
    \end{tabular}
    }
    \vspace{-8pt}
    \caption{
    Evaluation results for MaskVCT and the baseline methods on the LibriTTS-R test-clean dataset. 
    }
    \label{tab:results_main}
        \vspace{-2pt}
\end{table*}

\begin{table*}[ht]
    \centering
        \resizebox{0.94 \linewidth}{!}{%
    \begin{tabular}{l|c c |c c | c c| c c}
    \hline
    Model & Accented Source & Source$\rightarrow$Target
    & UTMOS ($\uparrow$) & Q-MOS ($\uparrow$) & S-SIM ($\uparrow$) & SS-MOS ($\uparrow$) & A-SIM ($\uparrow$) & AS-MOS ($\uparrow$)   \\
    \hline
    GT (L2) 
        & \cmark
        & $-$
        & \bf{3.01$\pm$0.06} 
            & \bf{3.49$\pm$0.24} 
        & \bf{0.940}  
            & \bf{4.70$\pm$0.13} 
        & \bf{0.519}  
            & \bf{4.51$\pm$0.16} \\
    \hline
    FACodec 
        & \xmark
            & Libri$\rightarrow$L2 
        &2.87$\pm$0.08 
            & 2.92$\pm$0.28
        & 0.768 
            & \underline{3.05$\pm$0.23} 
        & \underline{0.356}   
            & 3.21$\pm$0.25  \\ 

    MaskGCT-S2A 
    & \xmark
            & Libri$\rightarrow$L2 
    &\bf{3.21}$\pm$0.07 
        &  \bf{3.75$\pm$0.21} 
    & \underline{0.770}   
        & 2.89$\pm$0.27
    & 0.345 
        &  \bf{3.39$\pm$0.22}  \\
    FreeVC 
    & \xmark
            & Libri$\rightarrow$L2 
    &2.79$\pm$0.06 
        & \underline{3.65$\pm$0.21} 
    & 0.764 
        & 3.01$\pm$0.25  
    & 0.312  
        & 3.00$\pm$0.25  \\
    \bf{MaskVCT-Spk}
    & \xmark
            & Libri$\rightarrow$L2 
    &   \underline{3.15$\pm$0.08}
        & 3.44$\pm$0.21 
    & \bf{0.790} 
        &  \bf{3.23$\pm$0.27}
    &  \bf{0.362}  
        & \underline{3.33$\pm$0.26} \\
    \hline
    FACodec 
    & \cmark 
            & L2$\rightarrow$L2 
    & 2.68$\pm$0.08 
        & 3.03$\pm$0.27 
    &\bf{0.891}  
        &  3.70$\pm$0.29 
    & 0.385 
        & \bf{2.54$\pm$0.28} \\
    MaskGCT-S2A 
    & \cmark 
            & L2$\rightarrow$L2 
    & \bf{3.13}$\pm$0.05 
        &  \bf{3.35$\pm$0.29} 
    & \underline{0.890} 
        & \underline{3.82$\pm$0.25} 
    & \underline{0.395}   
        & 2.30$\pm$0.28\\
    FreeVC
    & \cmark 
            & L2$\rightarrow$L2 
    & 2.65$\pm$0.06 
        & \underline{3.28$\pm$0.25}     
    &0.868 
        & 3.35$\pm$0.25 
    & 0.345   
        &  2.29$\pm$0.23 \\
    \bf{MaskVCT-Spk}
    & \cmark 
            & L2$\rightarrow$L2 
    &  \underline{3.10$\pm$0.06} 
        & \underline{3.28$\pm$0.24}
    & {0.868} 
        & \bf{3.98$\pm$0.24} 
    &   \bf{0.406} 
        & \underline{2.48$\pm$0.26} \\
    \hline
    \end{tabular}
    }
    \vspace{-8pt}
    \caption{Accent Conversion Results, the audios are converted from \textbf{Libri}TTS-R or \textbf{L2}-ARCTIC to L2-ARCTIC.}
        \vspace{-8pt}
    \label{tab:accent}
    
\end{table*}

\subsection{Inference Setup}\label{section:exp_inference}
During every iteration, we select the top probability positions by Gumbel-Softmax \cite{gumbel} following the number of cosine mask schedules.
For each chosen position, we sample token indices to unmask using top-$k$ ($k=35$) followed by top-$p$ ($p=0.9$).
We mainly evaluate the total $N=64$ iterations with per-codebook steps [40,16, 2,1,1,1,1,1,1].

Since MaskVCT supports multiple modes, we evaluated a wide range of ablation studies for the large search space of CFG weights to identify the optimal CFG weights for each condition.
From these experiments, we propose two primary modes: \textbf{MaskVCT-All} and \textbf{MaskVCT-Spk}.
MaskVCT-All is conditioned by continuous linguistic features with CFG weights $(\omega_\text{all}, \omega_\text{spk}, \omega_\text{ling})=(1.5,\,1.0,\,1.0)$ including pitch-conditioned weights, it achieves superior intelligibility and source pitch-following but sacrifices some speaker similarity.
MaskVCT-Spk is conditioned by quantized linguistic features with 
CFG weights $(\omega_\text{all}, \omega_\text{spk}, \omega_\text{ling})=(0,\,2.0,\,0.5)$ omitting pitch conditioning yields the highest speaker similarity, though intelligibility is reduced.
For the accent conversion, we found that pitch-conditioned generation was affected by the L2-ARCTIC dataset's ambient noise, so we only tested MaskVCT-Spk with $(\omega_\text{all}, \omega_\text{spk}, \omega_\text{ling})=(0,2.5,0.5)$.

\subsection{Metrics}

We assess intelligibility with WER/CER from Whisper large-v3 \cite{whisper}, speaker similarity (S-SIM) as the cosine between WavLM speaker embeddings of the converted utterance and the target prompt \cite{wavlm}, pitch tracking via the F0 Pearson correlation (FPC), and automatic quality via UTMOS v2 \cite{utmosv2}.
Note that FPC is condition-dependent: higher is better for pitch-conditioned models, whereas for pitch-unconditioned models, whose goal is generating pitch from a prompt, a lower FPC can be preferable.
For the accent conversion, we measure accent similarity (A-SIM) by CommonAccent \cite{commonaccent}.
For the subjective measures, we collect human-evaluated mean opinion scores (MOS) from 1 to 5 for quality (Q-MOS), speaker similarity (SS-MOS).
Especially for the accent conversion test, we also collect accent similarity (AS-MOS).
For all MOS, we also show 95\% confidence intervals (CI).

\section{Results}
\label{sec:results}
\vspace{-3pt}

Table \ref{tab:results_main} presents results on the LibriTTS-R test-clean subset. 
While FACodec achieves the lowest WER and CER and the highest speaker similarity among baselines, it yields low quality metrics.
MaskGCT-S2A attains the highest UTMOS and Q-MOS but has higher WER and CER, and lower speaker similarity.
FreeVC offers the highest FPC among baselines for pitch tracking, but compromises on UTMOS and speaker similarity.
Our MaskVCT supports flexible trade-offs via CFG weight tuning.
Under the CFG-All setting, the model achieved the second-highest FPC in all cases, demonstrating precise source-pitch tracking and delivering the best intelligibility among masked models, albeit with some reduction in speaker similarity. 
In contrast, the CFG-Spk configuration prioritizes speaker fidelity, attaining the top speaker similarity of 0.895 and the lowest FPC of 0.157, enabling generation that diverges from the source pitch.
In subjective MOS evaluations, MaskVCT variants rank second in overall quality but lead in SS-MOS with 3.69, surpassing the next-best, MaskGCT with 3.47.
These results underscore MaskVCT’s ability to balance intelligibility and pitch controllability (CFG-All) against speaker similarity (CFG-Spk).

In the Libri$\rightarrow$L2, MaskGCT-S2A achieves the highest quality in both the objective and subjective metrics.
However, MaskVCT-Spk leads in speaker similarity and accent similarity and also attains the top SS-MOS and a competitive second in AS-MOS.
Although MaskVCT’s UTMOS is slightly lower than the MaskGCT-S2A, it has overlapped in CI and remains substantially more natural than others.
All systems except MaskVCT tend to preserve the original LibriTTS's American or British accent when converting to accented speech.
In the L2$\rightarrow$L2, MaskGCT-S2A again tops UTMOS with MaskVCT-Spk close behind, and MaskVCT-Spk attains the highest SS-MOS and matches baselines on AS-MOS. Overall, MaskVCT-Spk offers the strongest speaker and accent fidelity while maintaining competitive listening quality.

\vspace{-4pt}
\section{Conclusion}
\vspace{-4pt}

We present MaskVCT, a zero-shot voice conversion system that delivers unprecedented multi‑factor controllability through multiple classifier‑free guidances.
Unlike conventional VC methods, which are limited to a single conditioning setup, such as continuous vs. quantized linguistic features or pitch‑conditioned vs. unconditioned, MaskVCT integrates diverse conditions in a single model.  
This controllability and the state-of-the-art speaker similarity are enabled by the syllabic representation, which leaks less pitch information with the lowest FPC results, while other linguistic features exhibit greater leakage.
However, despite their advantages, both continuous and quantized syllabic representations can degrade intelligibility by causing misreadings, as they are not trained to reconstruct original speech. 
Moreover, because the syllabic quantizer is determined by K-means clustering \cite{syllablelm,sylber}, misreadings caused by incorrect syllable mappings cannot be easily corrected.
In future work, we plan to address this limitation by training our own quantized representation using vector quantization with masked model training, aiming to preserve the benefits of the syllable representations while reducing intelligibility errors.

\vspace{-4pt}
\footnotesize\section{Acknowledgement}
\vspace{-4pt}
\footnotesize This work was supported by the Office of the Director of National Intelligence (ODNI), Intelligence Advanced Research Projects Activity (IARPA), via the ARTS Program under contract D2023-2308110001. The views and conclusions contained herein are those of the authors and should not be interpreted as necessarily representing the official policies, either expressed or implied, of ODNI, IARPA, or the U.S. Government. The U.S. Government is authorized to reproduce and distribute reprints for governmental purposes notwithstanding any copyright annotation therein.

\vfill\pagebreak\newpage
\vfill\pagebreak
\newpage
\bibliographystyle{IEEEbib_abbrev}
{\footnotesize
\bibliography{main}}

\end{document}